\documentclass[12pt]{article}
\usepackage[utf8]{inputenc}
\usepackage[margin=1in]{geometry}
\usepackage{authblk}
\usepackage{amsmath}
\usepackage{amssymb}
\usepackage{hyperref}
\usepackage{float}
\usepackage{physics}
\usepackage{tikz}
\usetikzlibrary{decorations.pathmorphing}
\usetikzlibrary{arrows.meta}
\usepackage{fbb}
\usepackage{subcaption}
\usepackage[font=small]{caption}

\title{CFTs Blueshift Tensor Fluctuations Universally}
\author[1]{Matthew Baumgart}
\author[2]{Jonathan J. Heckman}
\author[1,3]{Logan Thomas}

\affil[1]{\textit{Department of Physics, Arizona State University, Tempe, AZ 85287}}
\affil[2]{\textit{Department of Physics and Astronomy, University of Pennsylvania, Philadelphia, PA 19104}}
\affil[3]{\textit{Beyond: Center for Fundamental Concepts in Science,
Arizona State University, Tempe, AZ 85287}}

\date{}

\newcommand{\beq}{\begin{equation}}
\newcommand{\eeq}{\end{equation}}
\newcommand{\ml}{\mathcal{L}}
\newcommand{\mo}{\mathcal{O}}
\newcommand{\nn}{\nonumber}
\newcommand{\eps}{\epsilon}

\begin{document}

\maketitle

\begin{abstract}
The strong constraints of conformal symmetry cause any nearly-conformal sector to blueshift tensor fluctuations in cosmology.
Hidden sectors with approximate conformal symmetry, which may be quite large, are a well-motivated extension of physics
beyond the Standard Models of particle physics and cosmology. They can therefore lead to a detectable shift in the tensor tilt for next-generation CMB and gravitational wave experiments. We compute the leading-order contribution to the in-in graviton two-point function from virtual loops in such sectors to demonstrate this universal effect. In units where a single conformally-coupled scalar is 1, limits from Stage-IV CMB experiments could bound the size of this extra sector to be smaller than $\sim\!\!10^{15}$.  This would be sufficient to rule out $N$-Naturalness as a complete resolution of the hierarchy problem.
\end{abstract}

\section{Introduction \label{sec:intro}}

Extra sectors beyond those of the Standard Model are
well-motivated both from a bottom-up and top-down perspective. In bottom-up terms, there are a wide variety of
possible dark matter scenarios. From a top-down perspective, string compactifications
typically come with a large number of extra sectors beyond those required to build the visible Universe.

Precisely because they are hidden, searching for experimental evidence of such sectors is clearly challenging.
A common phenomenological assumption is that dark matter interacts in some way with visible matter, but {\it a priori},
it might equally be the case that dark matter only interacts gravitationally. In the context of models
with many extra sectors (such as those coming from string theory), there is also no reason that such
sectors actually participate at all in the thermal history of the Universe, and in some scenarios,
the appearance of a large number of extra sectors can provide novel solutions to various hierarchy problems \cite{Arkani-Hamed:2016rle}.
The model-dependent nature of any given scenario thus makes its challenging to extract general lessons,
let alone possible experimental constraints.

Corrections to gravitational correlation functions provide an excellent opportunity to
extract potentially model-independent constraints on such sectors.
Indeed, the leading order coupling of fluctuations in the metric $h_{\mu \nu}$ to
stress energy $T_{\mu \nu}$ is universal:
\begin{equation}
\mathcal{L} \supset h_{\mu \nu} T^{\mu \nu}.
\end{equation}
In particular, the leading order loop contributions
to graviton in-in correlation functions from an extra sector amount to evaluating
in-in correlation functions of stress energy tensors in the extra sector.

Such correlation functions are directly observable. For example, corrections to
the graviton two-point function encode tensor mode fluctuations in inflationary cosmology.
More formally, such contributions also show up in the determination of the wave function of the Universe \cite{Maldacena:2011mk}.
Given all of this, it is natural to study how various extra sectors contribute to such correlation functions.  Some examples of these calculations can be found in the literature, including minimally-coupled scalars \cite{Adshead:2009cb,delRio:2018vrj}, and massless and massive fermions \cite{delRio:2018vrj,Feng:2012jm,Tan:2019czo}.\footnote{Like us, ref.~\cite{Frob:2012ui} even computed the modification of a conformal field to the graviton two-point function.  They only reported a result for exact de Sitter, and thus did not obtain a cosmological prediction.  As we will explain in Section \ref{sec:results}, there is actually no observable modification to the power spectrum in pure dS.}  Ref.~\cite{delRio:2018vrj} even conjectured a general formula for any number of approximately free massless scalars, fermions, and gauge bosons.  Unfortunately, this literature is confused in its current state.  Many of the papers disagree both with each other and, as we shall see, the answer demanded by conformal symmetry.  Clearing this up is one of the central points of this work.

Our aim in this paper will be to carry out this calculation
for extra sectors which are well-approximated by
conformal field theories (CFTs). CFTs do not introduce any additional mass scales,
and specify fixed points of renormalization group flow in a quantum field theory
with mass scales. As such, they are ubiquitous, and perturbations from a fixed point
provide a way to systematically move away from conformality and consider more general
QFT extra sectors. There are by now many examples of four-dimensional CFTs which run the gamut
from weakly-coupled free field theories, to strongly coupled non-Lagrangian systems.

In spite of this wide range of possibilities, conformal symmetry is enough to constrain the short distance behavior of the
stress energy tensor. The (convergent) operator product expansion for two stress tensors in a $D$-dimensional CFT is:
\begin{equation}
T_{\mu \nu}(x) T_{\rho \sigma}(0) = c \frac{I_{\mu \nu \rho \sigma}(x)}{x^{2D}}+ ...,
\label{eq:ttcorr}
\end{equation}
where $I_{\mu \nu \rho \sigma}(x)$ is a specific four-index tensor (see {\it e.g.}~\cite{Osborn:1993cr})
and $c > 0$ is a conformal anomaly / central charge of the CFT which can be viewed as counting the degrees of freedom
in the CFT.\footnote{In 4D CFTs, this is the Weyl density anomaly, which is distinct from the Euler density anomaly $a$,
as seen in the expression $\langle T_\mu^\mu\rangle = \frac{c}{16\pi^2}C_{\mu\nu\rho\sigma}C^{\mu\nu\rho\sigma} - \frac{a}{16\pi^2}E$.
For further discussion, see \cite{Duff:1993wm} and references therein.}
For example, in units of a single conformally-coupled real scalar $c_{\mathrm{scalar}}$, the
central charges for a Dirac fermion, a vector boson, and the $\mathcal{N} = 4$ vector multiplet of $U(N)$
Super Yang-Mills theory are:
\begin{align}
c_{\mathrm{Dirac}} &= 6 \, c_{\mathrm{scalar}} \nn \\
c_{\mathrm{vector}} &= 12 \, c_{\mathrm{scalar}} \nn \\
c_{\mathrm{\mathcal{N} = 4}} &= 27 N^2 \, c_{\mathrm{scalar}}.
\label{eq:ccharges}
\end{align}
Importantly, there are many examples of strongly coupled non-Lagrangian
CFTs where it is nevertheless possible to extract the value of $c_{\mathrm{extra}}$, the central charge of the hidden sector.\footnote{
Many calculable examples come from superconformal field theories (SCFTs), where the value of the conformal anomalies $a$ and $c$
can be obtained by calculating the relevant 't Hooft anomalies associated with the infrared R-symmetry \cite{Anselmi:1997am, Anselmi:1997ys}.
This in turn involves an application of the principle of $a$-maximization \cite{Intriligator:2003jj}. Anomalies of many
strongly coupled 4D SCFTs can also be extracted from the anomaly polynomial of a higher-dimensional 6D superconformal field theory
compactified on a Riemann surface using various generalizations of \cite{Benini:2009mz}. For a review
of methods used to compute the anomaly polynomial of 6D SCFTs, see e.g. \cite{Heckman:2018jxk} and references therein.}
In particular, this means that the relevant contribution to tensor fluctuations
from a conformal hidden sector can be boiled down to a single \textit{calculable} number.

Given its importance as a cosmological observable, it is perhaps surprising that only a few papers have attempted to calculate
such loop corrections, though as discussed, the results to be found in the literature are often contradictory, with differing signs, as well as
precise numerical values.
For us, all the different choices of a hidden sector only depend on $c_{\mathrm{extra}}$,
which for a CFT always has a fixed sign. Precisely because our answer is universal, it suffices to calculate the explicit form
of the correction to in-in two-point functions in the case of a single conformally-coupled scalar:
\begin{equation}
\label{confoscalar}
\mathcal{L}_{\phi} = \frac{1}{2} (\partial \phi)^2 - \frac{\xi}{2} \phi^{2} R,
\end{equation}
where $\xi = 1/6$ for the conformal case. At a technical level, this is somewhat simpler to carry out than the case of a general
CFT stress tensor, but universality ensures that our result applies for \textit{all} CFTs. Indeed, some care is required to extract the overall answer, because we are computing an in-in correlator in a cosmological background as opposed to the more conventional case of an in-out correlator as one would entertain in an S-matrix calculation.

Similar to the minimally-coupled scalar contribution computed in \cite{delRio:2018vrj}, we find that in single-field slow-roll inflation,
the contribution to tensor fluctuation power spectrum goes as:
\beq
\Delta \mathcal{P}_\gamma = +\eps_* \frac{2H_k^2}{\pi^2 M_{\rm pl}^2} \frac{N_{\rm CFT}}{16\pi^2} \frac{H_*^2}{M_{\rm pl}^2} \left[\frac{1}{15}\log \left( \frac{k}{k_*} \right) \right]\,,
\label{eq:tfpsintro}
\eeq
where $H_k$ is the Hubble parameter when the mode $k$ crosses the horizon, $\epsilon_*$ is the slow roll parameter and $H_*$ is the Hubble parameter, both of which are evaluated when the ``pivot scale,'' $k_*$, crosses the horizon.  To leading order in $\eps_*$, the relation is $H_k = H_* (H_*/k)^{\eps_*}$.  Even though, Eq.~\ref{eq:tfpsintro} and similar shifts to the power spectrum below are already at $\mo(\eps_*)$, we keep the overall $H_k^2$ from the graviton mode functions as it will allow us to easily factor out the contribution present at tree level.  Compared to the minimally-coupled scalar, the only difference in our case is the dimensionless factor, $1/15$.  We recall that the tensor power spectrum is defined as
\beq
\frac{k^3}{2\pi^2} \left \langle \gamma^{\mathbf{k}\,s}_{ij} (\eta) \, \gamma^{\mathbf{k^\prime}\,s}_{ij} (\eta) \right \rangle \equiv (2\pi)^3 \, \delta^{(3)}({\bf k} + {\bf k^\prime}) \, \mathcal{P}_\gamma,
\label{eq:tfpsdef}
\eeq
and the $\Delta \mathcal{P}_\gamma$ in equation (\ref{eq:tfpsintro}) is just the leading contribution beyond tree level.  The parameter $N_{\rm CFT} = c_{\mathrm{extra}} / c_{\mathrm{scalar}}$ counts the number of degrees of freedom in the CFT in units of the conformally-coupled scalar.
To best constrain $N_{\rm CFT}$ for a near-conformal sector that lacks non-gravitational interactions with the Standard Model, we need to observe primordial tensor fluctuations.\footnote{In section \ref{sec:pheno}, we discuss the possibility of large $N_{\rm CFT}$ running afoul of various ``strong-coupling bounds.''  If $r$ were limited to values much smaller than its present limit, and bounds on $n_t$ were not similarly strong, one may need to resort to a more theoretical limit of this type.  However, as we describe, such determinations are not sharp, or rely on speculative claims about black hole physics.}  Such a discovery is typically parametrized by
\beq
r \equiv \frac{\mathcal{P}_\gamma}{\mathcal{P}_\zeta},
\eeq
where $\mathcal{P}_\zeta$ is the usual curvature fluctuation power spectrum.  The other experimental quantity of interest is the running of $\mathcal{P}_\gamma$ with $k$, given by the {\it tensor tilt},
\beq
n_t \equiv \frac{d \log \mathcal{P}_\gamma}{d \log k}.
\label{eq:ntdef}
\eeq
In section \ref{sec:pheno}, we discuss current limits and near-future prospects for these parameters.
Our result in equation (\ref{eq:tfpsintro}) gives:
\beq
N_{\rm CFT} < 6.8\times10^{15}\left(\frac{0.01}{r}\right)^2 \left( \frac{n_t}{0.2} \right),
\label{eq:introbound}
\eeq
under the assumption $n_t \gg r$, as we expect if a large hidden sector is significantly modifying slow-roll inflation.  We note that our reference values for $r$ and $n_t$ in equation (\ref{eq:introbound}) sit in the regime which is expected to be probed by Stage-IV CMB experiments starting later this decade \cite{Wu:2014hta,Huang:2015gca}.

In F-theory compactifications, the number of D3-branes can often be of order 1,000\textendash 10,000 ({\it e.g.}~\cite{Klemm:1996ts,
Denef:2008wq, DelZotto:2016fju}), and there can be many additional extra sectors as well \cite{Taylor:2015xtz}.
Crudely then, the contribution to $N_{\rm CFT}$ could in principle be on the order of $10 N^2 \sim 10^{9}$.  Thus, ruling out these scenarios will require observation of $r$ near its present bound, and much stronger limits on $n_t$ than what CMB observations alone are projected to provide.  A more useful constraint is given on $N$-Naturalness as a complete explanation of the hierarchy problem \cite{Arkani-Hamed:2016rle}.  This requires $\sim 10^{16}$ copies of sectors like the Standard Model, which has (including right-handed neutrinos) $N_{\rm CFT} = 292$, treating it as approximately conformal.  Thus, the whole model has an effective $N_{\rm CFT} \sim 10^{18}$.  Even given the current limit on $n_t < 2.54$ for positive $n_t$ \cite{Planck2018}, detection of $r \sim 10^{-2}$ would rule out this explanation of the weak scale.

Our analysis makes manifest that all the contributions from the CFT sector,
be they weakly coupled bosons / fermions, or more strongly coupled composite objects all make
the \textit{same sign} contribution to tensor mode fluctuations, and it is a blueshifting of the spectrum.\footnote{This is a manifestation of the Le Ch\^{a}telier principle: slow roll inflation introduces a red-tilting of the power spectrum and the response of the scale invariant CFT is to counteract this by introducing a blue-tilt.}
Deviations from conformality provide one way to potentially modify the resulting phenomenology.
For example, we can also contemplate non-conformal couplings by changing the value of $\xi$ in equation (\ref{confoscalar}). It turns
out that a positive contribution to $\Delta \mathcal{P}_{T}$ requires working in the window centered on the conformal
case of $\xi = 1/6$, and outside this window, $\Delta \mathcal{P}_{T}$ is negative. Importantly, this latter case includes
the popular choice of a ``minimally-coupled scalar'' with $\xi = 0$. This, and other related issues suggest that low values of $\xi < 1/6$ might be problematic, but we leave a full treatment to future investigations.\footnote{For example, when $\xi = 1/6$, we can interpret the CFT as being placed on a compact $S^3$, and heated up to a temperature set by the de Sitter temperature. Taking $\xi < 1/6$ would then specify an effectively tachyonic contribution to the thermal Green's function.}

The rest of this paper is organized as follows. We begin in section \ref{sec:cftgrav} by presenting a general discussion of couplings between a CFT and gravity. In particular, we explain the sense in which we can extract a universal contribution
to the in-in graviton two-point function from in-in CFT correlation functions. In section \ref{sec:results} we turn to the core computation,
which amounts to calculating the explicit form of this correlation function for a conformally-coupled scalar.  We do this in two different ways; first we do an explicit loop calculation.  Then we take the well-known gravitational one-loop effective action and process it into a determination of the tensor power spectrum and tilt, finding a consistency check for the latter with the diagrammatic result.  In section \ref{sec:pheno} we turn to the phenomenological
implications of our calculation, highlighting that current and future bounds on tensor perturbations constrain the content of
extra sectors. Finally, in section \ref{sec:disc} we discuss the consequences of possible extensions beyond CFTs, and mention an interesting theory that gives zero correction to the tensor tilt at one loop. Some additional details on the various calculations are included in an Appendix.

\section{CFT/Gravity Interactions \label{sec:cftgrav}}

We begin with a basic review of coupling matter to gravitational fluctuations. Then on general grounds, we show why the effect of CFTs on the cosmological observable of tensor tilt is universal, scaling only with central charge.

We use the ADM formalism to decompose the metric, following \cite{Maldacena03}.  This gives a graviton field, $\gamma_{ij}$, that only has spacelike components.  At leading order in slow-roll, we will only need the tensor fluctuations of the metric \cite{delRio:2018vrj}.\footnote{Ref.~\cite{delRio:2018vrj} showed this for a minimally coupled scalar, but it just followed from expanding the coupling of gravity to $T_{\mu \nu}$ for the matter sector, and is thus generic.}  We will work in a transverse-traceless (TT) gauge.  For our cosmological case of interest, we can further specify the background geometry as de Sitter (dS).  In section \ref{sec:results} and Appendix \ref{app:loop}, we detail how our result is corrected by the slow-roll modifications to dS.  However, this is most simply done by straightforward modifications to the one-loop-corrected graviton propagator in pure dS as detailed in \cite{delRio:2018vrj,Senatore:2009cf}.  Therefore, it is easiest to begin with the metric
\begin{equation}
g_{ij} = a^2(\eta) \left(-\delta_{ij} + \gamma_{ij} \right),
\end{equation}
which for $\gamma_{ij} = 0$ is just the spacelike portion of dS in FRW-coordinates.  The function $a(\eta) = -1/(H\eta)$ is the usual scale factor, and $\eta = -e^{-Ht}/H$ is the conformal time.  The action of the graviton, $\gamma_{ij}$, at quadratic order is
\begin{equation}
    S_\gamma = \frac{M_{\rm pl}{}^2}{8}\int d^4 x \, \frac{1}{(H\eta)^2}\, \left[ \partial_\eta \gamma_{ij}\partial_\eta \gamma_{ij} - \, \partial_l \gamma_{ij}\partial_l \gamma_{ij} \right].
    \label{eq:gravitonaction}
\end{equation}
In general, the action of a spectator field on a curved background can be expanded in terms of the background metric. At linear order, this gives a universal coupling between the graviton and the stress tensor of the spectator fields,
\begin{align}
\mathcal{S}[\chi,\bar{g}_{\mu \nu}+h_{\mu\nu}] &= \mathcal{S}[\chi,\bar{g}_{\mu \nu}] + \int d^4 x \, h_{\mu \nu} \frac{ \delta \mathcal{S}[\chi,g_{\mu\nu}]}{ \delta g_{\mu \nu}} \Bigr|_{g_{\mu \nu }=\bar{g}_{\mu \nu }}+\mathcal{O}(h^2) \nn \\
&= \mathcal{S}[\chi,\bar{g}_{\mu \nu}] + \int d^4 x \frac{\sqrt{-\bar{g}}}{2} h_{\mu \nu} \, T^{\chi \, \mu \nu} [\bar{g}_{\mu \nu}]+\mathcal{O}(h^2), \nn \\
&= \mathcal{S}[\chi,\bar{g}_{\mu \nu}] + \int d^4 x \frac{a(\eta)^2}{2} \gamma_{ij} \, T^\chi_{ij} [\bar{g}_{\mu \nu}]+\mathcal{O}(\gamma^2),
\label{eq:gravmat}
\end{align}
where in the last line we go to our de Sitter case of interest, with $h_{ij}=a^2(\eta)\gamma_{ij}$.\footnote{Contraction with all lowered indices implies contraction using $\delta_{ij}$ instead of the metric.}

\begin{figure}
\centering
    \begin{tikzpicture}
        \draw[black, ultra thick, decorate, decoration=snake] ( 1, 0) -- ( 3, 0);
	\draw (2, 0.3) node {$-\mathbf{k}$};
        \draw[-{Stealth[scale length=2,scale width=3]}] (2,0) -- (2.01,0);
        \draw[black, ultra thick, decorate, decoration=snake] (-1, 0) -- (-3, 0);
	\draw (-2, 0.3) node {$\mathbf{k}$};
        \draw[-{Stealth[scale length=2,scale width=3]}] (-2,0) -- (-2.01,0);
        \draw[black, ultra thick] (0, 0) circle (1);
        \draw[-{Stealth[scale=2]}] (0.2,1) -- (0.21,1);
        \draw[-{Stealth[scale=2]}] (0.2,-1) -- (0.21,-1);
	\draw (0, 1.3) node {$\mathbf{p}$};
	\draw (0, -0.6) node {$-\mathbf{p} - \mathbf{k}$};
    \end{tikzpicture}
    \caption{The one-loop correction to the graviton propagator due to a hidden conformal scalar.}
    \label{fig:LoopDiagram}
\end{figure}
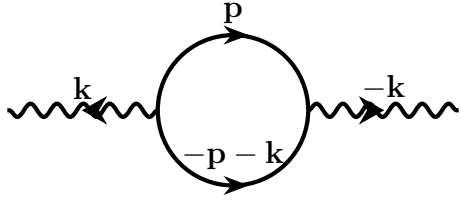

We can already see why a universal CFT modification, scaling only with central charge, $c$, to graviton propagation may be in the offing.  At second order in $H_{\rm int}$, the graviton two-point function will get a one-loop correction from the $\gamma_{ij} T^{ij}$ vertex in equation (\ref{eq:gravmat}) ({\it cf.}~figure (\ref{fig:LoopDiagram})).  This correction will have the form $\left \langle \gamma \, \gamma \, \gamma \, \gamma \, T \, T \right \rangle = \left \langle \gamma \, \gamma \right \rangle \left \langle \gamma \, \gamma \right \rangle \left \langle TT \right \rangle$, dropping indices for clarity.  Since $\left \langle TT \right \rangle$ is uniquely determined by conformal symmetry up to the central charge of the CFT (equation \eqref{eq:ttcorr}), we thus have a universal effect.

One must ask though, if this modification is the entirety of the CFT's effect on the tensor two-point function at one loop.  There are a few possible concerns, which we address in turn.  Firstly, having coupled the CFT to gravity, we are necessarily violating conformal symmetry.  This will ultimately cause $\left \langle TT \right \rangle$ to deviate from its pure-CFT form.  However, even though the conformal sector can generically involve strong dynamics, as long as we take the scale of inflation and the scales probed by our experiment to be $\ll M_{\rm pl}$, then corrections will be of $\mo((H,k)^2/M_{\rm pl}^2)$, where $k$ is some external momentum.

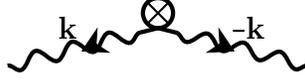
\begin{figure}
\centering
    \begin{tikzpicture}
        \draw[black, ultra thick, decorate, decoration={snake, segment length = 5mm}] ( 2, -0.7) -- (0, -0.2);
	\draw (1.2, -0.2) node {-$\mathbf{k}$};
        \draw[-{Stealth[scale length=2,scale width=3]}] (1,-0.5) -- (1.02,-0.51);
        \draw[black, ultra thick, decorate, decoration={snake, segment length = 5mm, amplitude = -1mm}] (-2, -0.7) -- (0, -0.2);
	\draw (-1.2, -0.2) node {$\mathbf{k}$};
        \draw[-{Stealth[scale length=2,scale width=3]}] (-1,-0.5) -- (-1.02,-0.51);
        \draw[black, ultra thick] (0, 0) circle (0.2);
        \draw[black, thick] (0.1414, 0.1414) -- (-0.1414, -0.1414);
        \draw[black, thick] (0.1414, -0.1414) -- (-0.1414, 0.1414);
    \end{tikzpicture}
    \caption{The $h^2$ vertex cannot contribute to the tensor tilt, as there is no external momentum dependence in the loop integral.}
    \label{fig:TwoGravitonDiagram}
\end{figure}

Another potential issue is the unspecified $\gamma^2$ vertex in equation (\ref{eq:gravmat}).  This allows the graviton to talk to the CFT through something other than its stress-energy tensor, threatening the universal correction.  Unlike the corrections just discussed, it can contribute at one loop and is unsuppressed by further powers of $M_{\rm pl}$.  We see though, that it cannot contribute to the tensor tilt since both gravitons participate in the local interaction it describes.  Whatever effect it has cannot modify the momentum dependence of the graviton propagator.  Any loop corrections due to the CFT will necessarily be independent of the graviton momentum ({\it cf.}~figure \ref{fig:TwoGravitonDiagram}).  Therefore, our tensor tilt cosmological observable will receive a correction from CFTs coming entirely from the convolution of the graviton propagators with $\left \langle TT \right \rangle$ as given by figure (\ref{fig:LoopDiagram}), and thus will depend only on the central charge of that sector.

Before proceeding to an explicit calculation, we address one final possibility of a strong violation of universality.  The interactions of equation (\ref{eq:gravmat}) are those of Einstein-Hilbert gravity coupled to matter.  Since we are already willing to work in the realm of quantum gravity and compute $M_{\rm pl}$-suppressed corrections, there may be further interactions arising at the Planck scale, which do not appear in the low-energy effective field theory one gets from quantizing the fluctuations about de Sitter in general relativity.  Implicitly though, we take the scale of inflation, as given by $H$, to be small compared to $M_{\rm pl}$ (otherwise our calculation is certainly incomplete).  We then simply take the standard coupling of long-distance gravity to matter via the latter's stress-energy tensor.  One is welcome to posit a modification to GR that would affect the physics of inflation while respecting current experimental constraints.  Meeting this challenge is beyond the scope of this work.  Given the intriguing success of Starobinsky inflation \cite{Starobinsky:1979ty,Starobinsky:1980te,Vilenkin:1985md} in satisfying the observational results of cosmology \cite{Planck2018}, it would be interesting to redo the analysis for that scenario.

\section{Conformally-Coupled Scalar \& Fermion Corrections \label{sec:results}}

As we showed in section \ref{sec:cftgrav}, the leading correction to the graviton two-point function from a CFT affects the cosmologically-observable tensor tilt comes entirely from convolving graviton propagators with the stress-energy two point function, $\left \langle T T \right \rangle$.  We proceed to do the calculation for two specific examples, a conformally-coupled scalar (CCS) and a massless Dirac fermion.  The ratio of the corrections is a factor of $+6$, just as dictated by conformal symmetry and the ratio of the central charges ({\it cf.}~equation \eqref{eq:ccharges}).  As mentioned in section \ref{sec:intro}, we then have the result for an arbitrary CFT, with central charge $c_{\rm extra}$, by scaling the single-CCS correction by $N_{\rm CFT} = c_{\rm extra}/c_{\rm scalar}$.

The structure of the scalar and fermion calculations is identical, and for both we follow the technique laid out in \cite{delRio:2018vrj,Senatore:2009cf}, which incorporates the effects of slow-roll as well, which are crucial for an observable effect.  The one-loop correction to the graviton two-point functions scales like $\eps$, the slow-roll parameter measured at some pivot scale, as found for the minimally coupled scalar in \cite{delRio:2018vrj}.  To begin though, we start in pure de Sitter and compute
\beq
\left \langle \gamma^{\mathbf{k}\,s}_{ij} (\eta) \, \gamma^{\mathbf{-k}\,s^\prime}_{ij} (\eta) \right \rangle = \left \langle \left( \overline T e^{+i \int d \eta^\prime \, H_{\rm int} (\eta^\prime)} \right)  \gamma^{\mathbf{k}\,s}_{ij} (\eta) \, \gamma^{\mathbf{-k}\,s^\prime}_{ij} (\eta)  \; \left( T e^{-i \int d \eta^{\prime\prime} \, H_{\rm int} (\eta^{\prime\prime})} \right) \right \rangle
\label{eq:initcalc}
\eeq
We work to second order in $H_{\rm int}$, as given in equation (\ref{eq:gravmat}).  Furthermore, as argued in section \ref{sec:cftgrav}, for the graviton momentum, $k$, -dependent corrections that will affect the tensor tilt, we only need those contributions with two insertions of the $\gamma_{ij} T^{ij}$ vertex.  We will refer to this correction as $\left \langle \gamma^{\mathbf{k}\,s}_{ij} (\eta) \gamma^{\mathbf{-k}\,s^\prime}_{ij} (\eta) \right \rangle_2$.  As it stands, equation (\ref{eq:initcalc}) is in the usual, interaction picture, in-in form, with the state in the expectation value the standard Bunch-Davies vacuum \cite{Bunch:1978yq}.  As a technical simplification, we use the trick of deforming the time contour to run from $\eta - i\, \infty$ to $\eta + i\, \infty$, and the entire calculation is anti-time-ordered in the Euclidean, or imaginary time component (AETO perturbation theory), as in \cite{Green:2020txs}.   Thus, in practice we compute
\beq
\left \langle \gamma^{\mathbf{k}\,s}_{ij} (\eta) \, \gamma^{\mathbf{-k}\,s^\prime}_{ij} (\eta) \right \rangle = \left \langle \overline T_E  \left( \gamma^{\mathbf{k}\,s}_{ij} (\eta) \, \gamma^{\mathbf{-k}\,s^\prime}_{ij} (\eta)  \; e^{\int_{-\infty}^{\infty} d \eta^\prime \, H_{\rm int} (\eta + i\, \eta^\prime)} \right)\right \rangle.
\label{eq:aetocalc}
\eeq
What follows is an overview of the one-loop correction to the graviton two-point function, $\left \langle \gamma^{\mathbf{k}\,s}_{ij} (\eta) \gamma^{\mathbf{-k}\,s^\prime}_{ij} (\eta) \right \rangle_2$, from a conformally-coupled scalar and a free fermion.  The ``2'' subscript again indicates that we only determine the contribution that shifts the observable tensor tilt, $n_t$ ({\it cf.}~equations \eqref{eq:tfpsdef} and \eqref{eq:ntdef}).  We thus ignore contributions involving the two-graviton vertex, as discussed at the end of section \ref{sec:cftgrav}.  We provide calculational details in Appendix \ref{app:loop}.

\subsection{Conformally-Coupled Scalar \label{subsec:ccs}}

As argued in section \ref{sec:cftgrav}, the leading modifications to the tensor tilt come only from the linear perturbation in $\gamma_{ij}$.  For general matter in a de Sitter background, using the ADM formalism, this coupling is
\beq
\ml_{\rm int} = \int d^4 x \frac{a(\eta)^2}{2} \gamma_{ij} \, T_{ij},
\label{eq:genint}
\eeq
where we note that all indices are lowered and are thus contracted with $\delta_{kl}$.  Specifying that our extra sector is a conformally-coupled scalar (CCS), we have the following scalar Lagrangian,
\beq
S_\phi = \int d^4x \, \frac{1}{2(H\eta)^2} \left[ (\partial_\eta \phi)^2 - (\partial_i \phi)^2 - \frac{\xi}{(H\eta)^2} R \, \phi^2 \right],
\eeq
where $\xi = \frac 1 6$ for the conformal case, and $R = 12H^2$ for dS.  The stress tensor of a conformally-coupled scalar field on a de Sitter background is
\begin{equation}
    T_{\mu\nu} = \nabla_\mu\phi\nabla_\nu\phi-\frac{1}{2}g_{\mu\nu}g^{\alpha\beta}\nabla_\alpha\phi\nabla_\beta\phi+\frac{1}{6}\left[g_{\mu\nu}\Box-\nabla_\mu\nabla_\nu+G_{\mu\nu}\right]\phi^2\,,
\end{equation}
where $G_{\mu\nu}$ is the Einstein tensor in de Sitter space. The tracelessness of the graviton polarization tensor, combined with the fact that $\epsilon_{0\mu}=\epsilon_{\mu0}=0$, implies that much of the stress tensor will not contribute to the two-point function. Therefore, we will define the \textit{interacting part} of the stress tensor, $T_{\mu\nu}^{\text{int}}$, by discarding all terms which are manifestly proportional to the (spatial part of the) metric, and ignore $0^{\rm th}$ components of tensors. In dS, the Einstein tensor is proportional to the metric and the Christoffel symbol is proportional to the spatial part of the metric,
\begin{equation}
    \Gamma_{ij}^\lambda = g_{ij}g^{\lambda\rho}\partial_\rho\ln[H\eta]\,.
\end{equation}
Therefore, we may write the interacting part of the stress tensor as
\begin{equation}
    T_{ij}^{\text{int}} = \frac{2}{3}\partial_i\phi \, \partial_j\phi - \frac{1}{3} \phi \, \partial_i\partial_j\phi\,.
\label{eq:trel}
\end{equation}
Plugging equation (\ref{eq:trel}) into equation (\ref{eq:genint}), we thus get the following interaction,\footnote{One may wonder whether coupling of the curvature field, $\zeta$, could provide an additional limit.  In fact, any interaction between $\phi$ and $\zeta$ will be suppressed relative to the coupling to gravitons by a factor of $\eta$ \cite{delRio:2018vrj}.  Thus, the effect on the scalar tilt is suppressed by $\eps^2$, and this will give a weaker bound despite the stronger constraints on $n_s$.}
\beq
\ml_{\rm int} = \int d^4 x \frac{a(\eta)^2}{2} \gamma_{ij} \, \left( \frac{2}{3}\partial_i\phi \, \partial_j\phi - \frac{1}{3} \phi \, \partial_i\partial_j\phi \right).
\label{eq:phiint}
\eeq
It is straightforward to proceed by putting the interaction of equation (\ref{eq:phiint}) into the interaction-picture determination of $\left \langle \gamma^{\mathbf{k}\,s}_{ij} (\eta) \gamma^{\mathbf{-k}\,s^\prime}_{ij} (\eta) \right \rangle_2$ given by equation (\ref{eq:initcalc}).  We provide the details in Appendix \ref{app:loop}.  The steps closely follow the similar calculations of \cite{delRio:2018vrj,Senatore:2009cf}, although we compute in anti-Euclidean-time-ordered (AETO) perturbation theory, which provides useful technical simplifications.

It is worth mentioning here that like \cite{delRio:2018vrj,Senatore:2009cf}, we compute with dimensional regularization.  Just as in \cite{delRio:2018vrj}, we fully renormalize our one-loop result at late times with counterterms from dimension-6 operators.  However, one may worry that the logarithms arising from this RG that mixes operators of different dimensions are unphysical, just as mass-dependent schemes in flat space can introduce spurious breakdowns in power counting.\footnote{We thank Daniel Green for bringing this concern to our attention.}  It is nevertheless true that in a nonrenormalizable QFT like quantum gravity, we do expect the generation of increasingly-irrelevant operators in perturbation theory.  Rather than litigate the matter directly, our resolution is to perform an independent calculation that sidesteps dimreg entirely.  In section \ref{subsec:cck}, we compute the tensor power spectrum and tilt directly from the graviton one-loop effective action, finding agreement with the dimreg results reported below.

In total, taking the late-time limit in a pure dS background, the 1-loop correction to the graviton propagator from $N(=N_{\rm CFT})$ conformally-coupled scalars is\footnote{Strictly, this is the one-loop correction that can, in principle, affect the tensor tilt, and thus does not include contributions from the $\gamma^2$ vertex, as explained in section \ref{sec:cftgrav}.  Also, we have subtracted the divergent constant to give a renormalized result.}
\begin{equation}
\frac{k^3}{2\pi^2}\left \langle \gamma^{\mathbf{k}\,s}_{ij} (0) \gamma^{\mathbf{k^\prime} s^\prime}_{ij} (0) \right \rangle_2 = -\frac{N_{\rm CFT}}{8\pi^2}\left(\frac{H}{M_{\rm pl}}\right)^4 \delta^{s s^\prime} \delta^{(3)}(\mathbf{k}+\mathbf{k^\prime})\left[\frac{4\pi}{15}\log \left( \frac{H}{\mu} \right) \right]\,.
\label{eq:puredscorr}
\end{equation}
Thus, in exact de Sitter spacetime, this loop correction is a constant. In fact, by making the seemingly judicious choice of renormalization scale, $\mu = H$, we can remove it entirely.\footnote{Although this can be removed {\it at late times}, in general the loop correction does have time dependence and cannot be removed by a local counterterm~\cite{Senatore:2009cf}.} One way to understand the absence of running is the self-similarity of pure dS.  The spacetime possesses an exact rescaling symmetry under $a \rightarrow \lambda \, a,\, x \rightarrow x/\lambda,\, k \rightarrow \lambda \, k$.  This is one way to see that the naive result with $\log(k/\mu)$ instead of $\log(H/\mu)$ must be incorrect.  Thus, we cannot get a $k$-dependent correction.\footnote{One might wish to avoid the rescaling argument by replacing the comoving momentum, $k$, with a physical momentum, $k/a$, in the log.  However, our results will implicitly take the $\eta \rightarrow 0$ late-time limit, so we cannot have explicit time dependence in that case.}  This correction can be directly compared to equation (B.39) in \cite{delRio:2018vrj}.  Since no other relative factors will enter in converting the result of equation (\ref{eq:puredscorr}) to an inflationary observable, we see that a conformally-coupled scalar gives -1/9$^{\rm th}$ the correction of a minimally coupled scalar.\footnote{As a cross check on our AETO calculational methods, we also used them to compute the one-loop correction due to a minimally coupled scalar, reproducing the result of \cite{delRio:2018vrj}.}  We will discuss this ratio, and how it follows from the graviton one-loop effective action, further in section \ref{subsec:cck}.

If we instead consider a Universe with a small slow-roll parameter $\epsilon$, then we can do an analogous computation as in Ref.~\cite{delRio:2018vrj,Senatore:2009cf}, which replaces the $\log(H/\mu)$ with $\log(H_k/H_*) = -\epsilon_* \log(k/k_*)$, where we have chosen $\mu = H_*$, the Hubble scale when the experimental pivot scale (around which we measure our cosmological observables) exits the horizon, and $H_k$ is $H$ when the $k$ mode exits.  We can see the momentum dependence more clearly in the version that explicitly depends on $k$, with $k_*$ as the pivot scale, and $\eps_*$ is the slow-roll parameter determined when $k_*$ crosses the horizon.  Thus, the one-loop graviton propagator correction in inflation (to leading power in $\eps_*$) is
\begin{equation}
\frac{k^3}{2\pi^2}\left \langle \gamma^{\mathbf{k}\,s}_{ij} (0) \gamma^{\mathbf{k^\prime} s^\prime}_{ij} (0) \right \rangle_2 = +\eps_* \frac{N_{\rm CFT}}{8\pi^2}\frac{H_k^2H_*^2}{M_{\rm pl}^4} \delta^{s s^\prime} \delta^{(3)}(\mathbf{k}+\mathbf{k^\prime})\left[\frac{4\pi}{15}\log \left( \frac{k}{k_*} \right) \right]\,.
\end{equation}
We can convert this to the tensor power spectrum ({\it cf.}~equation \eqref{eq:tfpsdef}), getting
\beq
\Delta \mathcal{P}_\gamma = +\eps_* \frac{2H_k^2}{\pi^2 M_{\rm pl}^2} \frac{N_{\rm CFT}}{16\pi^2} \frac{H_*^2}{M_{\rm pl}^2} \left[\frac{1}{15}\log \left( \frac{k}{k_*} \right) \right]\,.
\label{eq:explcorr}
\eeq
Summing the geometric series of diagrams generated by this loop gives
\beq
\mathcal{P}_\gamma = \frac{2H_k^2}{\pi^2 M_{\rm pl}^2} \left[ 1 - \eps_* \frac{N_{\rm CFT}}{16\pi^2} \frac{H_*^2}{M_{\rm pl}^2}\frac{1}{15} \log \left( \frac{k}{k_*} \right) \right]^{-1}.
\label{eq:fulltfps}
\eeq
Although we have done this calculation as an explicit one scalar loop Feynman diagram, since the gravitons are just coupling to $T_{\mu \nu}$ ({\it cf.}~equation \eqref{eq:genint}), the correction $\Delta \mathcal{P}_\gamma$ just depends on the two-point function $\langle TT \rangle$, which is universal for CFTs up to the positive central charge.  Thus, taking $N_{\rm CFT} = c_{\rm extra}/c_{\rm scalar}$, where $c_{\rm extra}$ is the central charge of the CFT of interest, equations (\ref{eq:explcorr}) and (\ref{eq:fulltfps}) hold for {\it all conformal field theories}.

\subsection{Massless Fermion \label{subsec:fermion}}

As a nontrivial check on equation (\ref{eq:explcorr}), we should compute the one-loop correction to the tensor power spectrum in a different CFT.  A free, massless Dirac fermion provides a natural candidate.  As further motivation for this particular theory, such a calculation has been done multiple times in the literature \cite{delRio:2018vrj,Feng:2012jm,Tan:2019czo}, with none of the papers matching the CFT prediction ($N_{\rm CFT} = 6$ for a single Dirac fermion), nor each other.  Ref.~\cite{delRio:2018vrj} disagrees only by a minus sign though, and does report a correction from a free U(1) gauge boson consistent with equation (\ref{eq:explcorr}).  We therefore just do the computation from scratch.

The fermion couples to the graviton through its stress-energy tensor (equation \eqref{eq:genint}), giving the following interaction,
\begin{equation}
\ml_{\rm int} = \frac{i}{4}\int\,d^4x\,a^3(\eta) \left[\bar{\psi}\, \gamma_{(i}\partial_{j)}\psi-\partial_{(i}\bar{\psi}\gamma_{j)}\psi\right] \gamma_{ij}\,.
\label{eq:fermint}
\end{equation}
Computing the one-loop correction to the graviton propagator from the coupling to a massless fermion proceeds just as the correction from the CCS does.  We thus need only account for the superficial differences in their respective interaction terms, as we would expect.  Calculational details are given in Appendix \ref{app:loop}.  We find the correction due to a single massless Dirac fermion is $+6$ times that from a single CCS, just as expected from our argument based on conformal symmetry in section \ref{sec:cftgrav}, and the ratio of central charges in equation (\ref{eq:ccharges}).  Concretely, we get
\beq
\Delta \mathcal{P}^{\rm Dirac}_\gamma = +\eps_* \frac{2H_k^2}{\pi^2 M_{\rm pl}^2} \frac{1}{16\pi^2} \frac{H_*^2}{M_{\rm pl}^2} \left[\frac{2}{5}\log \left( \frac{k}{k_*} \right) \right]\,.
\label{eq:fermcorr}
\eeq
Thus, for an extra CFT sector with $n_D$ Dirac fermions, we can use the results of equations (\ref{eq:explcorr}) and (\ref{eq:fulltfps}) with $N_{\rm CFT} = 6n_D$.

\subsection{Cross Check}
\label{subsec:cck}

As a cross check on our result, we compare to that from the one-loop graviton effective action, determined in \cite{Christensen:1978gi,BirrellDavies,Vassilevich:2003xt},
\begin{align}
S_{\rm eff} &= -\frac{M_{\rm pl}^2}{2}\int\,d^4x\sqrt{-g}R - \frac{1}{5760\pi^2}\int\,d^4x\sqrt{-g}\left[a \, W_{\mu\nu\rho\sigma}W^{\mu\nu\rho\sigma}+b\, (R_{\mu\nu}R^{\mu\nu}-\frac{1}{3}R^2) + d\, R^2\right]\,, \nn \\
&= -\frac{M_{\rm pl}^2}{2}\int\,d^4x\sqrt{-g}R - \frac{1}{5760\pi^2}\int\,d^4x\sqrt{-g}\left[\left (a + \frac b 2 \right) W_{\mu\nu\rho\sigma}W^{\mu\nu\rho\sigma} + dR^2\right]\,,
\label{eq:weyleffaction}
\end{align}
where $W_{\mu\nu\rho\sigma}$ is the Weyl tensor. Looking at table \ref{tbl:loopeff} and equation \eqref{eq:weyleffaction}, we see that for CFTs, $d=0$, and so the deformation to Einstein-Hilbert is due entirely to the square of the Weyl tensor, $\sim W^2$.  Using the Gauss-Bonnet relation in 4 dimensions, we have
\begin{equation}
S_{\rm eff} = -\frac{M_{\rm pl}^2}{2}\int\,d^4x\sqrt{-g}R - \frac{1}{5760\pi^2}\int\,d^4x\sqrt{-g}\left[(b+2a)R_{\mu\nu}R^{\mu\nu}+\left(d-\frac{b+2a}{3}\right)R^2\right]\,.
\label{effaction}
\end{equation}
Ref.~\cite{Salvio:2017xul} found the result that in the presence of
\beq
\ml_{\rm gravity} \supset -\frac{1}{2f_2^2} \int d^4 x \, \sqrt{-g} \, W_{\mu\nu\rho\sigma}W^{\mu\nu\rho\sigma},
\eeq
the tensor power spectrum ({\it cf.} equation~\eqref{eq:fulltfps}) is modified by an overall factor,
\beq
\mathcal{P}_\gamma = \frac{2H^2}{\pi^2 M_{\rm pl}^2} \left[ \frac{1}{1 + \frac{4H^2}{f_2^2\, M_{\rm pl}^2}} \right].
\label{eq:eatfps}
\eeq
Looking at table \ref{tbl:loopeff}, we see that for the CCS, $f_2 = 8\sqrt{30} \pi$.  Plugging this in, along with the fact that $dH^2/d \log k = -H^2\, r/8$ in slow-roll inflation, gives the shift in $n_t \equiv d\, \log{\mathcal P_\gamma}/d\, \log k$,
\beq
\Delta n_t = +\frac{r \, H^2}{3840 \pi^2 M_{\rm pl}^2},
\label{eq:nteff}
\eeq
This matches the result in \eqref{eq:fulltfps}, up to $H^4/M_{\rm pl}^4$ corrections, since $\eps_* = r/16$ (see also equation~\eqref{eq:tiltcorr} below).  Furthermore, different CFTs just change the overall $W^2$ factor by $N_{\rm CFT}$; the agreement holds trivially for them, as well.  Thus, the computation passes a nontrivial cross check, and for this calculation, dimreg provided a legitimate method.

The coefficients $a$, $b$, and $d$ are determined by the field content which has been integrated out ({\it cf.}~table \ref{tbl:loopeff}).\footnote{In \cite{delRio:2018vrj}, the spin-$1/2$ factors differ from table (\ref{tbl:loopeff}) by a relative sign.  This would violate the prediction from conformal symmetry.  As we show in section (\ref{subsec:fermion}) and the Appendix, explicit calculation upholds the sign in table \ref{tbl:loopeff}. This error has been acknowledged to us\cite{SubodhEmail}.}
\begin{table}[tp]
\begin{center}
\begin{tabular}{|c|c|c|c|} \hline
Spin & a & b & d \\ \hline
0 & 1 & 1 & $90(\xi - 1/6)^2$ \\ \hline
1/2 & 7/2 & 11 & 0 \\ \hline
1 & -13 & 62 & 0 \\ \hline
\end{tabular}
\caption{Contributions to terms in one-loop graviton effective action (equation~\eqref{eq:weyleffaction})}
\label{tbl:loopeff}
\end{center}
\end{table}
For a CFT, the combinations of $a$, $b$, and $d$ coefficients in equation~\eqref{eq:weyleffaction} are
\begin{align}
b+2a &= 3N_{\rm CFT}\\
d-\frac{b+2a}{3} &= -N_{\rm CFT}\,.
\end{align}
Since de Sitter is a maximally symmetric spacetime, $R_{\mu \nu} = R g_{\mu \nu}/4$.  Thus, the net result of the one-loop effective action, as written in equation~\eqref{effaction} is to add an $R^2$ term, whether the theory is conformal or not.  Taking into account the nonzero $d$ in nonconformal cases, one can also determine their modification to $n_t$ from table \ref{tbl:loopeff}.  The ratio of general scalar theory's $\Delta n_t$ to that of a CCS ({\it cf.}~equation \eqref{eq:nteff}) is just given by ratio of their $R^2$ coefficients in equation~\eqref{effaction}, after replacing $R_{\mu \nu} \rightarrow R g_{\mu \nu}/4$

We note that even though the leading modification to $n_t$ is identical whether one uses Feynman diagrams \& dimreg, or the effective action, the modifications to the power spectrum, $\mathcal P_\gamma$, itself are naively different. This is of no consequence when $N_\text{CFT}$ is small, as the corrections are higher order in $H^2/M_\text{pl}^2$. In the limit where $N_{\text{CFT}}$ becomes large, the $W^2$ term in the effective action dominates, as the overall $f_2^{-2} \propto N_\text{CFT}$.  Thus, the predicted power spectrum becomes scale invariant.  This is unsurprising as the $W^2$ lagrangian is that of conformal gravity, which appears to dominate over Einstein-Hilbert.  We see a different behavior though, in our diagrammatic calculation of $P_\gamma$ (equation \ref{eq:fulltfps}).  The corrections are a perturbation  so long as the term proportional to $N_\text{CFT} \log(k/k_*) \ll 1$. It is therefore incumbent on us to determine which result to use for very large hidden sectors.

To frame the discussion to follow, it is helpful to recall a similar situation in particle physics involving Higgs boson decays to vector bosons.
The $h \rightarrow VV$ process is captured by a dimension five operator of the schematic form $\frac{h}{v} \mathrm{Tr} F^2$, in the obvious notation. Now, one can extract the leading order contribution to this higher dimension operator from states which pick up their mass from the Higgs vev $v$ by computing the one-loop beta function to the gauge coupling, with the mass of the corresponding threshold corrections controlled by $v$, {\it i.e.}:
\begin{equation}\label{runthatcoupling}
\frac{4 \pi}{g^2(\mu)} = \frac{4 \pi}{g^2(m)} + \frac{\delta b_{G}}{2 \pi} \, \mathrm{log}( \mu  / m),
\end{equation}
with $m = \lambda v$ the mass of the state and $\lambda$ the corresponding Yukawa coupling. Making the substitution $m \mapsto m + \lambda h$ and plugging in the corresponding value of $1 / g^2$ into the vector boson kinetic term one gets a leading order contribution to the corresponding decay process:
\begin{equation}
\delta b_{G} \mathrm{log} \left(1 + \frac{h}{v} \right) \mathrm{Tr} F^2,
\end{equation}
so expanding in powers of $h/v$ yields the desired dimension five operator. Said differently,
the size of the threshold correction from the corresponding heavy state determines the strength of the dimension five operator.

In fact, since the threshold correction measures the departure from conformal invariance, there is a corresponding mixed gauge / conformal symmetry anomaly which exactly captures this threshold correction \cite{Shifman:1978,Shifman:1979eb,Shifman:1986}. See also \cite{Heckman:2012nt}
for further discussion on this point in the context of supersymmetric theories with visible / hidden sector mixing.
There are also limitations to this approximation. For example, the contribution from the anomaly is one-loop exact, and does not capture more detailed information such as the mass and momentum dependence of the threshold states being integrated out. To some extent, this can be absorbed into the value of $4 \pi / g^2(m)$ appearing in equation (\ref{runthatcoupling}).

Our situation is quite analagous. Indeed, the term $\mathrm{Tr} F^2$ in our case is instead captured by
the Weyl$^2$ term, which is indeed just the kinetic term in a theory of conformal gravity. From this perspective, we are simply calculating a
threshold correction to the conformal gravity term as induced by a CFT loop correction. As far as the conformal anomaly goes, this contribution is one-loop exact, and amounts to completely integrating out the CFT Hilbert space from high momenta all the way to low momenta. This is, of course, not quite the same calculation as we performed via a direct diagrammatic computation (where the momentum dependence was retained at all stages).  To one-loop order, in the limit of vanishing momentum, when detailed momentum dependence becomes irrelevant, we can clearly interchange the two calculations. Beyond this leading order approximation, however, we will face the same subtleties as those already mentioned in the context of Higgs / Vector Boson couplings. Some of this can again be absorbed into a modified value of Newton's constant, much as we can modify the running of the gauge coupling in the Higgs physics example.

Observe that equation (\ref{eq:fulltfps} )contains results to all orders in perturbation theory, capturing the physics when $N_\text{CFT}$ is sufficiently large that two-loop and higher terms in the geometric series are numerically important.  Additionally, we obtained the effective action (equation (\ref{eq:weyleffaction})) by integrating out a massless CFT.  Since the correction to the tilt scales like $\sim N_\text{CFT} \log(k/k_*)$, the large $N_\text{CFT}$ and large momentum limit give the same result, and we should be wary of the nonlocal high-scale theory described with a large massless sector integrated out.  Reference \cite{Shifman:1986} details the modifications for a class of field theories necessary to obtain the Wilsonian effective action appropriate for computing observables in a low-energy EFT from the 1PI-effective action.  Nonetheless, the one-loop, 1PI effective action is describing the physics of the conformal anomaly, and thus for small $N_\text{CFT}$ (which corresponds to low-momentum), we can trust its determination of the overall factor, and use it as a cross-check of our diagrammatic results for those theories. A fully rigorous proof that our result is more accurate at large $N_{\text{CFT}}$ would require a full two-loop calculation, which is outside the scope of this work. That being said, all two-loop contributions not included in our bubble-chain diagrams are necessarily suppressed. With this in mind, in the following section, we shall assume our result in Eq.~\eqref{eq:fulltfps} is the correct result to use in the large-$N_\text{CFT}$ limit, and use it to study the phenomenology of hidden CFT sectors with large $N_{\text{CFT}}$.

Lastly, we reiterate our comments from the end of section \ref{sec:cftgrav}.  Looking at the form of equation \eqref{eq:weyleffaction}, it is natural to wonder what else might contribute to the one-loop gravity effective action.  We are necessarily working in a low-energy effective field theory of the full quantum gravity, and thus {\it something} has surely been integrated out above our cutoff scale.  In particular, it could contribute some $a^\prime,b^\prime$ that would give an overall modification to the $W^2$ term and change, or perhaps negate and reverse the effect of our CFT.  This is certainly a logical possibility, just like the presence of a large additional hidden sector of {\it minimally}-coupled scalars that could actually provide a net redshift.  Nonetheless, absent a mechanism for making it so, it would be a fine-tuning if the modifications to gravity coming from the deep UV were large enough to substantially shift $n_t$, but otherwise leave slow-roll inflation (or even GR-itself) untouched.  Thus, for the discussion of observable constraints in section \ref{sec:pheno} below, we take the hidden CFT as the only source of a large $n_t$ shift.

\section{Phenomenology \label{sec:pheno}}

At the pivot scale, the loop contribution vanishes, as constructed. The effect of the hidden CFT sector is therefore seen in the tilt of the tensor power spectrum, $n_t$.  The hidden conformal sector modifies this relationship. The observational definition of the tensor tilt is
\begin{equation}
n_t \equiv \frac{d \log\mathcal{P}_\gamma}{d \log k}\bigg{|}_{k=k_*}.
\end{equation}
In slow-roll inflation, without extra field content, there is a consistency condition relating $r$ and $n_t$ at lowest order in the slow roll parameter, $n_t = -r/8$, which comes from the scale-dependence of $H$ in the tree-level power spectrum ({\it cf.}~\cite{Baumann:2009ds} for discussion).
Using the definition of $n_t$ and our result Eq.~\eqref{eq:fulltfps}, we can determine how the consistency relation for the tilt is modified by the presence of the conformal sector,
\begin{equation}
n_t= -\frac{r}{8} \left(1 - \frac{N_{\rm CFT}}{960}A_s \, r\right),
\label{eq:tiltcorr}
\end{equation}
where $A_s \, r = \frac{2H_*^2}{\pi^2 M_{\rm pl}^2}$, and we have plugged in the slow-roll relation $\eps_* = r/16$.  We see from equation (\ref{eq:fulltfps}) that $A_s \, r$ is just the tree-level contribution to $\mathcal{P}_\gamma$. In the following, we use a pivot scale of $k_*=0.01\text{Mpc}^{-1}$.
Inputting the value for $A_s$ measured by \textit{Planck} ($A_s = 2.2 \times 10^{-9}$) in \cite{Planck2018}, equation (\ref{eq:tiltcorr}) shows that the contribution of the CFT becomes comparable to that from slow roll around $N_{\rm CFT} = 4.4\times10^{13}(\frac{0.01}{r})$.  Stage-IV CMB experiments such as the Simons Observatory, Lite-Bird, and others are slated to begin operations in the late 2020s.  They are projected to place 95\% CL limits on $|n_t| < 0.2 $ \cite{Wu:2014hta,Huang:2015gca}.  Since CFTs blueshift the power spectrum, we are interested in the upper limit, $n_{t\,\text{max}}$.  We find a limit on $N_{\rm CFT}$ for $n_{t\,\text{max}} \gg r$:\footnote{Using model-independent priors on the value of $r$ at two different pivot scales, the \textit{Planck} collaboration placed a bound on $n_t$, at a pivot scale of $k_*=0.01\, \text{Mpc}^{-1}$, within the ranges $-0.55<n_t<2.54$ at 95\% CL \cite{Planck2018}.  Using LIGO/VIRGO data, one can set more aggressive limits on $n_t$ (see \cite{Giare:2020vss} for an example), but our calculation is not reliable at the many-decades-shorter wavelengths probed by these gravity wave experiments.}
\begin{align}
N_{\rm CFT} &< 7.0\times10^{15}\left(\frac{0.01}{r}\right)^2 \left( \frac{n_{t\,\text{max}}}{0.2} \right), \nn \\
N_{\rm CFT} &< 3.7 \times 10^{14}, \quad (r=0.044,\,n_{t\,\text{max}} = 0.2).
\label{nbound}
\end{align}
\begin{figure}[t!]
\centering
\begin{subfigure}[t]{0.45\textwidth}
\includegraphics[width=\textwidth,height=0.3\textheight]{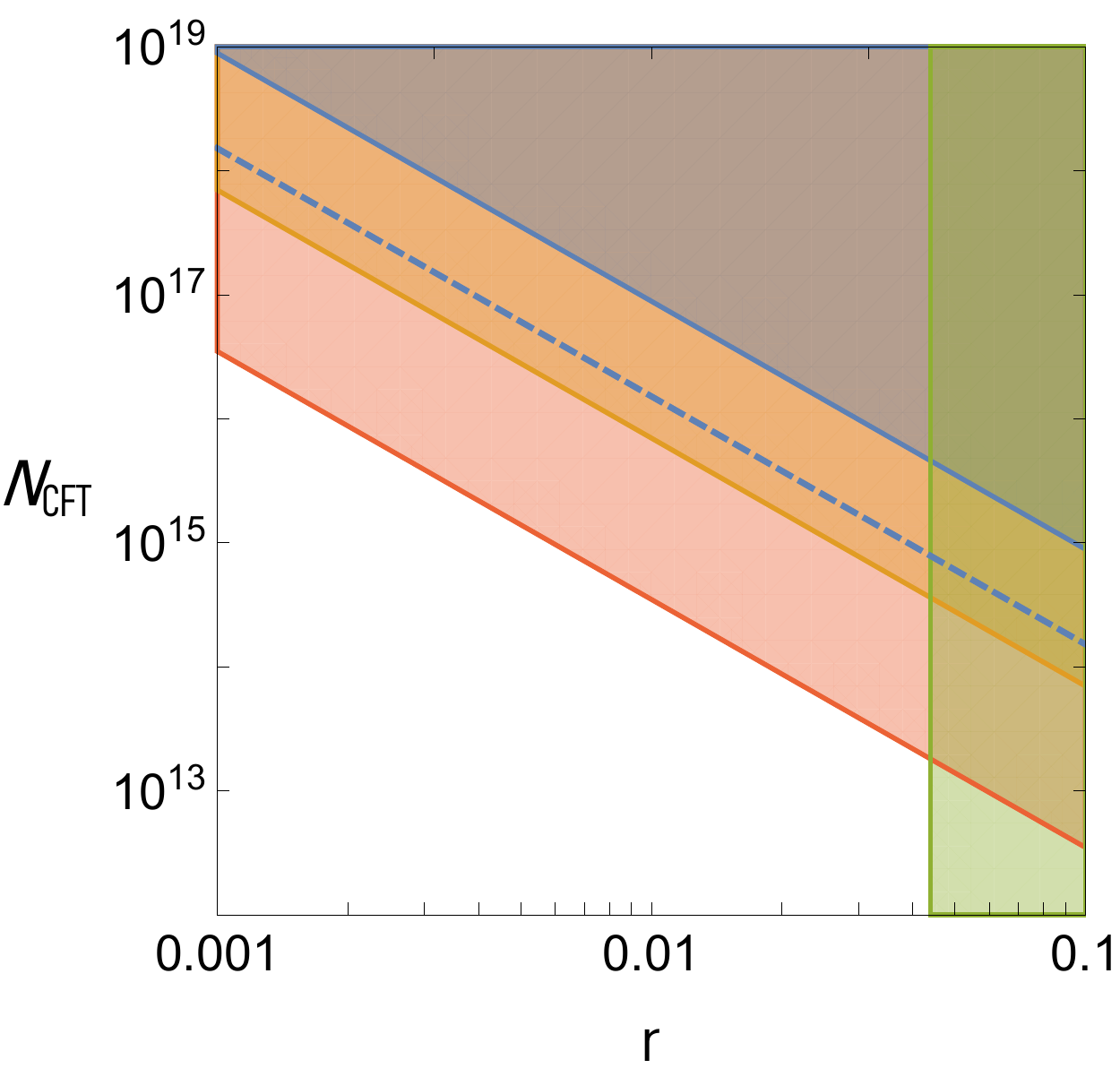}
\label{subfig:ExclusionPlotA}
\end{subfigure}
\hfill
\begin{subfigure}[t]{0.45\textwidth}
\includegraphics[width=\textwidth,height=0.3\textheight]{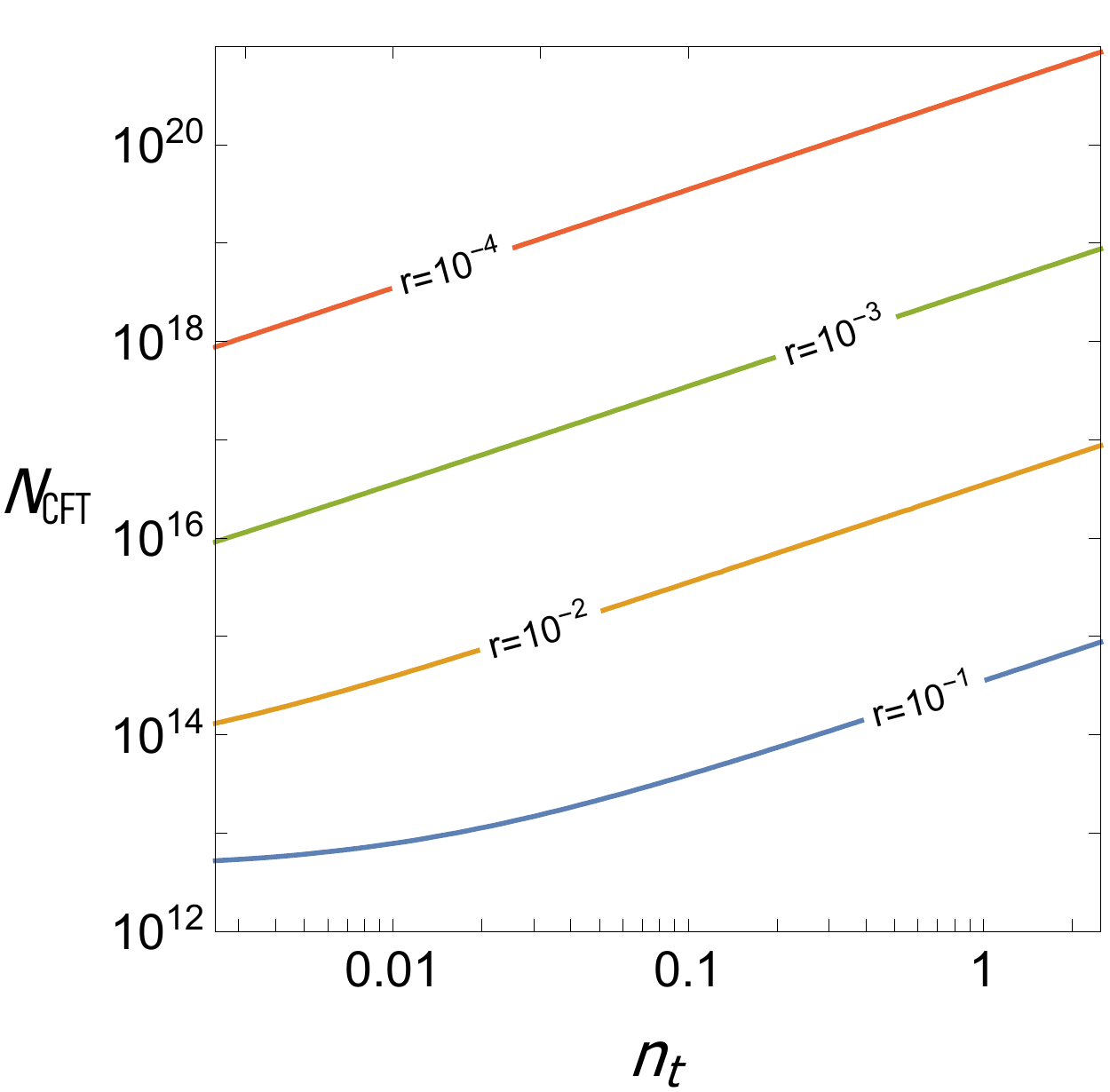}
\label{subfig:ExclusionPlotB}
\end{subfigure}
\caption{{\bf Left:} Constraints on $N_{\rm CFT}$ and $r$ based on current and projected bounds. The blue shaded region is excluded based on the current bound from Planck of $n_t<2.54$ \cite{Planck2018}. The orange shaded region is the excluded region given the projected bound of $n_t<0.2$ from Stage-IV CMB \cite{Wu:2014hta,Huang:2015gca}. The red shaded region is the excluded region given the projected bound $n_t<0.01$ from future space-based gravitational wave interferometers\cite{Friedman:2006zt,Campeti:2020xwn}. The green region shows the currently excluded values of $r$. The dashed blue line gives the values of $N_{\rm CFT}$ and $r$ for which the loop correction has a pole at $k=10k_*$.  {\bf Right:}  Predicted value of $N_{\rm CFT}$ from equation (\ref{eq:tiltcorr}), given different possible measurements of $r$ and $n_t$.}
\label{fig:lims}
\end{figure}
In the second line, to get the most constraining bound, we set $r$ equal to its current limit, found in a recent analysis of Planck and BICEP2/Keck data, $r<0.044$ \cite{Tristram:2020wbi}.  Stage-IV experiments are projected to bound $r \lesssim 10^{-3}$ absent a discovery \cite{Wu:2014hta,Huang:2015gca}.  Looking ahead to advanced space-based gravitational wave interferometers like BBO and DECIGO, bounds are projected to improve to $r < 10^{-6},\, |n_t| < 0.01$ \cite{Friedman:2006zt,Campeti:2020xwn}.  As long as $r$ is not too small, this limit on $n_t$ is sufficiently strong that the much shorter-frequency range probed by these instruments is still in the regime of perturbative validity for our calculation.  We show a summary of these current \& projected limits, along with the value of $N_{\rm CFT}$ that a particular combination of $r$ and $n_t$ would specify in figure (\ref{fig:lims}).

\begin{figure}
\centering
\begin{subfigure}{0.45\textwidth}
    \centering
    \begin{tikzpicture}
        \draw[black, ultra thick, decorate, decoration=snake] ( 1.3, 0) -- ( 2.5, 0) node[anchor=west] {\large $\sim N^2\frac{H^6}{M_{pl}{}^6}$};
        \draw[black, ultra thick] (-1.3, 0.1) -- (-0.5, 0.1);
        \draw[black, ultra thick] (-1.3,-0.1) -- (-0.5,-0.1);
        \draw[black, ultra thick, decorate, decoration=snake] (-0.5, 0) -- (0.5, 0);
        \draw[black, ultra thick] (1.3, 0.1) -- (0.5, 0.1);
        \draw[black, ultra thick] (1.3,-0.1) -- (0.5,-0.1);
        \draw[black, ultra thick, decorate, decoration=snake] (-1.3, 0) -- (-2.5, 0);
    \end{tikzpicture}
    \label{subfig:HigherLoopsA}
\end{subfigure}
\begin{subfigure}{0.45\textwidth}
    \centering
    \begin{tikzpicture}
        \draw[black, ultra thick, decorate, decoration=snake] (-3, 0) -- (-1.5, 0);
        \draw[black, ultra thick] (-1.5, 0.1) -- (1.5, 0.1);
        \draw[black, ultra thick, decorate, decoration={snake,segment length = 10.1pt}] (-1,0.1) arc (180:0:0.9);
        \draw[black, ultra thick] (-1.5,-0.1) -- (1.5,-0.1);
        \draw[black, ultra thick, decorate, decoration=snake] (1.5, 0) -- (3, 0);
        \draw[] (3,-0.9) node[anchor=west] {\large $\sim N\frac{H^6}{M_{pl}{}^6}$};
        \draw[black, ultra thick, decorate, decoration=snake] (3, -1) -- (0.25, -1);
        \draw[black, ultra thick] (0, -1) circle (0.25);
        \draw[black, ultra thick] ( 0.177, -1.177) -- (-0.177,-0.823);
        \draw[black, ultra thick] (-0.177, -1.177) -- ( 0.177,-0.823);
        \draw[black, ultra thick, decorate, decoration=snake] (-0.25,-1) arc (115:425:0.707);
        \draw[black, ultra thick, decorate, decoration=snake] (-0.25,-1) -- (-3,-1);
    \end{tikzpicture}
    \label{subfig:HigherLoopsB}
\end{subfigure}
    \caption{Of the two-loop diagrams contributing to the tensor power spectrum, the diagram constructed from two insertions of the stress tensor is parametrically larger than other possible diagrams for $N\gg1$. There are also diagrams involving scalar perturbations, but they are further suppressed by a factor of $\epsilon$. Double lines represent stress tensor insertions, and the $\bigotimes$ represents an insertion of the CFT operator that appears in the two-graviton vertex.}
    \label{fig:HigherLoops}
\end{figure}
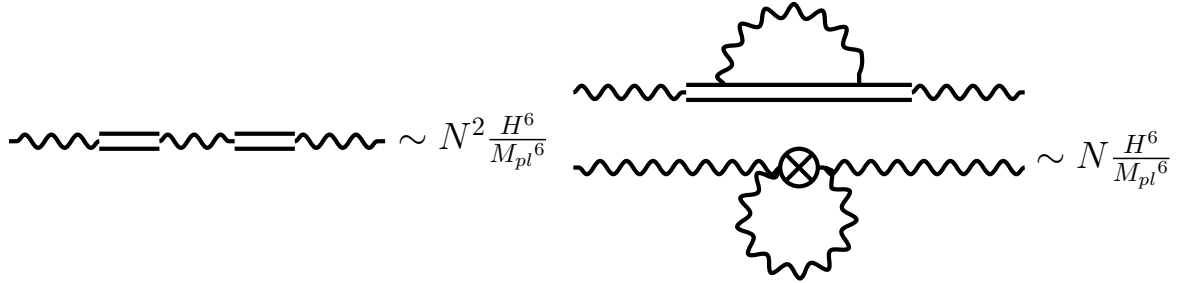

We can compare the bound above (equation \eqref{nbound}), based on current data and a systematic calculation, to tests of perturbative consistency. The strong coupling bound purports to set an upper bound on the number of particle species that may exist in a Universe where gravity can still be treated semi-classically\cite{Dvali:2007hz,Dvali:2007wp}.\footnote{The original bounds of \cite{Dvali:2007hz,Dvali:2007wp} arise from concerns over correct black hole evaporation times.  While interesting, these limits necessarily combine speculative assumptions about black holes at the scale of quantum gravity and assumptions about the emissivity of possibly exotic CFT matter.  We will not discuss them further.}  Broadly speaking, the claim is that this effective field theory for quantum gravity has a cutoff of
\beq
\Lambda \leq \frac{4\pi M_{\rm pl}}{\sqrt{N}},
\label{eq:scbound}
\eeq
where $N$ is some counting of the degrees of freedom.  Naively plugging in $N_{\rm CFT}$ from equation (\ref{nbound}) appears problematic as that would give $\Lambda < H$.  It is worth asking though, how equation (\ref{eq:scbound}) arises.

The parameter $M_{\rm pl}$ does not run in the familiar sense ({\it cf.}~ \cite{Donoghue:2019clr} for a concise overview of this point).  We thus need to choose a quantitative criteria to decide when quantum gravity effects have spoiled our perturbative calculation.  Ref.~\cite{delRio:2018vrj} claims a bound when the one-loop and tree-level contributions to equation (\ref{effaction}) become comparable.  However, there is nothing wrong {\it per se} with having one-loop effects comparable, or even larger than those at tree level.  This effect, known as having a ``large $k$-factor,'' is well known in particle physics \cite{Rubin:2010xp}.  The situation there is broadly similar to what we have here.  Observables at one loop become sensitive to other sectors which are not present at tree level.  Further radiative corrections beyond one loop can still be small.  Indeed, the Standard Model maintains perturbativity at LHC scales.  For our case of interest, drawing the diagrams that correct the graviton two-point function at two loops shows no new contributions that go like $N^2$ since gravity couples to the states of the CFT diagonally ({\it cf.}~figure (\ref{fig:HigherLoops}), we have of course the non-1PI graph already accounted for in the geometric series obtained from summing insertions of our 1PI loop).  Furthermore, our perturbative expansion parameter, $H^2/M_{\rm pl}^2 \sim (r/0.01) \times 10^{-11}$, serves to efficiently suppress the contributions from higher orders.

As a better test of strong coupling, we propose the following.  Summing the geometric series of one-loop insertions on the graviton propagator (the left diagram in figure \ref{fig:HigherLoops} is the second-order contribution in this series) gave equation~\eqref{eq:fulltfps}. We see that for $k$ sufficiently large compared to $k_*$, this will hit a singularity.  We can call this scale $k_{\rm pole}$.  Thus, a valid calculation should have $k_{\rm pole} \gg k_*$.  If $N_{\rm CFT}$ were sufficiently large so that $k_{\rm pole} \approx k_*$, this would not necessarily bound $N_{\rm CFT}$.  It would just preclude setting the bound from the one-loop calculation that gives equation (\ref{eq:fulltfps}).  We see in figure (\ref{fig:lims}), that Stage-IV CMB experiments will push into the region where $k_{\rm pole} > 10\times k_*$, providing for a perturbatively-consistent calculation.

\section{Discussion \label{sec:disc}}

In the previous sections we showed, both through general considerations as
well as in specific free field models, that a CFT\ always contributes with a
universal sign to the tensor power spectrum, causing it to blueshift. Of
course, the assumption of conformal symmetry in an extra sector is something
of an idealization, so it is interesting to ask about how these results are
modified for more general QFTs.

In many cases of interest, we can realize a QFT\ with mass scales as a
relevant deformation of a CFT. Close to the fixed point, this is specified
as a deformation of the Lagrangian density of the form:%
\begin{equation}
\mathcal{L}\rightarrow \mathcal{L}_{CFT}+\lambda O,
\end{equation}%
where $O$ is some spin zero operator of scaling dimension $\Delta <4$. The
parameter $\lambda $ implicitly specifies a characteristic mass scale $m\sim
\lambda ^{1/\Delta }$. For example, a massless scalar can be deformed by
adding a mass term $\frac{1}{2}m^{2}\phi ^{2}$. We can evaluate the
corrections to the stress energy tensor from such contributions, and this
will involve (model-dependent)\ correlation functions involving the operator
$O$. Such an approximation is valid provided we work at energy scales where $%
\lambda O$ is a small perturbation.

As we approach the deep infrared where the perturbation is no longer small,
another fixed point (possibly trivial)\ will emerge, with a new value of the
conformal anomaly $c_{\text{IR}}$. In many cases of interest, $c_{\text{UV}%
}-c_{\text{IR}}>0$, though it is interesting to note that it is possible to
obtain a negative value \cite{Anselmi:1997ys}, a
feature which is distinct from $a_{\text{UV}}-a_{\text{IR}}$, which is
always positive \cite{Cardy1988, Komargodski:2011vj}. Using this, one can in
principle consider extra sectors where $c$ increases after a relevant
deformation. For such families of theories, then, the contribution to the
power spectrum along the flow will contain negative contributions, although
the fact that $c > 0$ at both the beginning and end of the flow would still suggest
an overall blueshift to the tensor fluctuation power spectrum.

This does, however, hint at the possibility that before reaching a new fixed
point, a non-conformal sector might be capable of generating an opposite-sign contribution to the power spectrum. At a practical level, one way to
parameterize such an effect is in terms of the proxy of a single scalar, but
with a more general choice of conformal coupling, namely taking $\xi \neq 1/6
$ in the effective field theory of a massless scalar:%
\begin{equation}
\mathcal{L} \supset - \frac{\xi}{2} \phi ^{2}R.
\end{equation}%
Indeed, we observe that at the level of the Hankel functions used to
parameterize the mode expansions of a scalar with a general choice of mass $m
$ and curvature coupling coupling $\xi$, the
contribution to the in-in two-point function can be packaged in terms of a
massless scalar with some shifted value of $\mathcal{\xi }$. This is in
accord with the fact that in a de Sitter background, the coupling $\phi ^{2}R
$ amounts to specifying a thermal mass for the scalar.

Restricting then, to the case of a massless scalar but with a general value
of $\xi $, we can use the heat kernel methods of \cite{Vassilevich:2003xt} to extract the leading order contribution
to the power spectrum. This results in a $\xi $ dependent value of $N(\xi )$, given by:\footnote{For a scalar of mass $m$ and non-conformal
coupling $\xi \phi ^{2}R$, the one-point function for the stress tensor, $\langle T^\mu_\mu \rangle$,
which is proportional to the one-loop graviton effective action, also
picks up a scheme-dependent contribution, $m^{2}\left\langle \phi \left( x\right) \phi (x)\right\rangle$
(see {\it e.g.}~Ch.~6 of \cite{BirrellDavies}). Clearly though, this vanishes in the massless limit.}
\begin{equation}
N(\xi )=-9-360\xi \left( \xi -\frac{1}{3}\right) ,
\end{equation}
a polynomial quadratic in $\xi $. We observe that in the conformally-coupled
case, $N = 1$, while in the \textquotedblleft minimally coupled
scalar\textquotedblright\ with $\xi =0$, $N(\xi )=-9$. A
curious feature of this result is that  there is only a small window of
values for which $N(\xi )$ is actually positive:%
\begin{equation}
N(\xi )>0\text{ \ \ for \ \ }-\frac{\sqrt{10}}{60}< \left( \xi -\frac{1%
}{6} \right) <\frac{\sqrt{10}}{60}.  \label{BrokenWindows}
\end{equation}%
We also note that $N(\xi )$ has a maximum precisely at the
conformal case of $\xi =1/6$.

From the perspective of a CFT\ calculation, taking $\xi <1/6$ would appear
to correspond to heating up the CFT\ to the de Sitter temperature, but with
a tachyonic mass. The case of $\xi >1/6$, however, corresponds to an
ordinary mass term, and as such, would appear to be perfectly valid. That
being said, to reach a large value of $\xi $ would appear to require
crossing through a value of $N(\xi )$ which vanishes: such a
contribution would be invisible to the tensor power spectrum! This alone suggests
something subtle (possibly pathological) may be happening outside of the
window defined by line (\ref{BrokenWindows}), though we leave a full
treatment for future work.

The detection of tensor fluctuations would also open up the possibility
of additionally observing tensor non-gaussianities ({\it e.g.}~\cite{Shiraishi:2019yux}).  This would
be particularly exciting in the presence of a large, hidden CFT sector, as it could
probe the structure of $\left \langle TTT \right \rangle $correlation functions. We remark that this correlator depends
on the Euler density conformal anomaly ({\it cf.}~\cite{Osborn:1993cr}),
which provides a related notion of ``counting degrees of freedom'' in a CFT.
The details of this, and the possible need to compute other CFT correlation
functions is left for future work.

Until Nature gives us experimental access to quantum gravitational fluctuations, we must allow for the possibility of hidden sectors with seemingly absurd size.  The result of this work (summarized in equation \eqref{eq:tiltcorr}) shows that CFTs and their perturbative and/or relevant deformations blueshift the tensor tilt.  Much attention to date for strongly blueshifted scenarios has come from heterodox theories like non-inflationary cosmology or structural modifications to spacetime ({\it cf.}~\cite{Calcagni:2020tvw} for several examples).\footnote{A more mundane model that gives a strong blueshift is axion-SU(2) inflation \cite{Dimastrogiovanni:2016fuu}.  Interestingly, the gravity waves in this scenario are strongly chiral, which gives it a possibility of being quickly disentangled from the more exotic scenarios, including ours.}  In the absence of other signals, the mere existence or a large hidden sector that otherwise leaves inflation and spacetime intact would seem a conservative, and arguably more likely explanation.  Thus, precisely measuring $n_t$ could provide a census for the entire Universe.  Were it to be found quite close to its prediction from single-field inflation ($n_t = -r/8$), we would learn at a glance that either 1) Nature has no enormous CFT-like hidden sectors, 2) or if it does they must be cancelled by similarly huge sectors that redshift $n_t$, like the minimally coupled scalar.  With the exception of the bizarre $\xi = 1/6 \pm \sqrt{10}/60$ possibility mentioned above, such a precise cancellation would demand an explanation, likely rendering the scenario implausible.  By contrast, measuring $n_t$ far removed from $-r/8$ would give strong evidence for the existence of an ocean of new states, with the sign of the shift characterizing the basic property of the new matter.

\section*{Acknowledgements}

We thank A. Saurabh for collaboration at an early stage of this work. We thank
P. Adshead, D. Green, T. Hartman, C. Keeler, J. Maldacena, M. Montero, S. Patil and D. Simmons-Duffin for helpful discussions and correspondence.
JJH thanks the ASU department of physics for kind hospitality during part of this work, as well as the
2021 Simons summer workshop at the Simons Center for Geometry and Physics
for kind hospitality during part of this work. MB and LT are supported by the DOE (HEP) Award DE-SC0019470.
The work of JJH is supported by the DOE (HEP) Award DE-SC0013528. The work of LT is also supported by the
Foundational Questions Institute (FQXi).

\appendix
\section{Loop Corrections to Tensor Tilt}
\label{app:loop}

\subsection{Conformally-Coupled Scalar}

The bulk of our calculational technique follows the method of \cite{delRio:2018vrj,Senatore:2009cf}, albeit with anti-Euclidean-time-ordered (AETO) perturbation theory, as in \cite{Green:2020txs}.

We can write the graviton field in momentum space as,
\begin{equation}
    \gamma^{\mathbf{k}\, s}_{ij} = \epsilon_{ij}^s(\mathbf{k}) \, \gamma^s_{\mathbf{k}}(\eta) \, a^s_{\mathbf{k}} + \epsilon_{ij}^{*\, s}(-\mathbf{k}) \, \gamma^{*\,s}{\mathbf{k}}(\eta) \, a^{\dag\, s}_{-\mathbf{k}} \,,
\end{equation}
with $\epsilon_{ii}^s=k_i\epsilon_{ij}^s=0$, $\epsilon_{ij}^s\epsilon_{ij}^{*s'}=4\delta_{ss'}$, and $\epsilon^{*s}(\mathbf{k})=\epsilon^s(-\mathbf{k})$.  We can see from the action, equation~\eqref{eq:gravitonaction}, that each helicity mode has effectively the same equation of motion as a massless scalar. We begin with fluctuations about a pure de Sitter geometry, and it is straightforward to deform our result to the case of slow-roll inflation.  The mode functions of the graviton in $d$ spatial dimensions are thus identical to those of a massless scalar, except for a division by $M_{\rm pl}$,
\begin{align}
    \gamma^{s}_\mathbf{k}(\eta) & =\frac{iH}{M_{\rm pl}\sqrt{2k^3}}(1+ik\eta)e^{-ik\eta} & (d=3) \nn \\
    &= -\frac{\sqrt{\pi}}{2} e^{i\pi \delta/4} \frac{H^{1+\delta/2}}{M_{\rm pl} \mu^{\delta/2}} \frac{(-k\eta)^{(3+\delta)/2}}{k^{(3+\delta)/2}}\, H^{(1)}_{(3+\delta)/2}(-k \eta) & (d=3+\delta)
\label{eq:gravmode}
\end{align}

We are interested in the Euclidean anti-time ordered propagator that shows up in our loop diagram, with one end at time $\eta$ and the other at time $\eta+i\eta_1$:
\begin{equation}
    \left\langle\bar{T}_E\left[\gamma^{\mathbf{k}\,s}_{ab}(\eta)\gamma^{-\mathbf{k}\,s}_{cd}(\eta+i\eta_1)\right]\right\rangle = \epsilon^s_{ab}(\mathbf{k})\epsilon^{*\,s}_{cd}(\mathbf{k})\frac{H^2}{M_{\rm pl}{}^2}\left(1+k^2\eta^2+ik^2\eta\eta_1+k|\eta_1|\right)\frac{e^{-k|\eta_1|}}{2k^3}\,.
\end{equation}
As for the conformal scalar, its mode functions are the same as in Minkowski space, but scaled by the conformal factor of the spacetime. This gives a Euclidean anti-time-ordered propagator
\begin{equation}
    \left\langle\bar{T}_E\left[\phi_{\mathbf{k}}(\eta+i\eta_1)\phi_{-\mathbf{k}}(\eta+i\eta_2)\right]\right\rangle = a^{-1}(\eta+i\eta_1)a^{-1}(\eta+i\eta_2)\frac{e^{-k|\eta_1-\eta_2|}}{2k}\,,
\end{equation}
with $a(\eta) = -1/H\eta$.  Using the interaction in equation (\ref{eq:phiint}), the loop diagram can then be written as
\begin{align}
\left \langle \gamma^{\mathbf{k}\,s}_{ij} (\eta) \gamma^{\mathbf{-k}\,s^\prime}_{ij} (\eta) \right \rangle_2
&= \delta^{s s^\prime}\int_{-\infty}^{\infty}\,d\eta_1\int_{-\infty}^{\infty}\,d\eta_2\int d^3p\, \left(2p_ap_bp_cp_d \right) \left(\frac{H^2}{M_{\rm pl}^2}\right)^2 \epsilon_{ab}^{*\, s}(\mathbf{k}) \, \epsilon_{cd}^{s^\prime}(\mathbf{k})  \nonumber\\
    &\hspace{.1in} \times \frac{e^{-k|\eta_1|}}{2k^3}\frac{e^{-k|\eta_2|}}{2k^3}(1+k^2\eta^2+ik^2\eta\, \eta_1+k|\eta_1|)(1+k^2\eta^2+ik^2\eta\, \eta_2+k|\eta_2|) \nn \\
    &\hspace{.1in} \times \frac{e^{-p|\eta_1-\eta_2|}}{2p}\frac{e^{-|k+p|\,|\eta_1-\eta_2|}}{2|k+p|},
\label{1loop}
\end{align}
where as discussed below equation (\ref{eq:aetocalc}), the ``2'' subscript on the LHS means we are only interested in the loop contribution that can give external graviton momentum dependence from the $T_{ij} \gamma_{ij}$ vertex in $\ml_{\rm int}$ ({\it cf.}~equation \eqref{eq:genint}).
It is simplest to compute the $\eta_{1,2}$ integrals first, resulting in
\begin{align}
\left \langle \gamma^{\mathbf{k}\,s}_{ij} (\eta) \gamma^{\mathbf{-k}\,s^\prime}_{ij} (\eta) \right \rangle_2 &=
   \left(\frac{H}{M_{\rm pl}}\right)^4\int d^3p\, \int\,d^3q\,\delta^3(\mathbf{q}+\mathbf{p}+\mathbf{k})\, p_a \, p_b \, p_c \, p_d \, \epsilon_{ab}^{*\, s}(\mathbf{k}) \, \epsilon_{cd}^{s^\prime}(\mathbf{k}) \nonumber \\ & \quad \times\frac{16k^3+29k^2(p+q) + 20k(p+q)^2 + 5(p+q)^3}{8k^7p\, q(k+p+q)^4}
\end{align}
where we have defined $\mathbf{q}=-\mathbf{k}-\mathbf{p}$, and taken the late-time limit. Further, $|p_ap_b\epsilon_{ab}|=p^2\sin^2\theta$, where $\theta$ is the angle between the momenta $\mathbf{p}$ and $\mathbf{k}$. The angular dependence can be written in terms of $p$, $q$, and $k$ using $q^2=p^2+k^2+2pk\cos\theta$.
\begin{align}
\left \langle \gamma^{\mathbf{k}\,s}_{ij} (\eta) \gamma^{\mathbf{-k}\,s^\prime}_{ij} (\eta) \right \rangle_2 &=
    \left(\frac{H}{M_{\rm pl}}\right)^4\frac{1}{k^6}\int d^3p\, \int\,d^3q\,\delta^3(\mathbf{q}+\mathbf{p}+\mathbf{k})\frac{(4p^2k^2-(q^2-p^2-k^2)^2)^2}{16k^4} \nn \\
    &\quad \times\frac{16k^3+29k^2(p+q) + 20k(p+q)^2 + 5(p+q)^3}{8k\, p\, q\, (k+p+q)^4}.
\label{eq:aftereta}
\end{align}
Taking
\beq
f(q,p,k) = \frac{(4p^2k^2-(q^2-p^2-k^2)^2)^2}{16k^4} \frac{16k^3+29k^2(p+q) + 20k(p+q)^2 + 5(p+q)^3}{8k\, p\, q\, (k+p+q)^4},
\eeq
for $d=3 + \delta$ spatial dimensions, by dimensional analysis, the above will evaluate to
\begin{align}
\int\,d^3p\int\,d^3q\,\delta^3(\mathbf{q}+\mathbf{p}+\mathbf{k})\,f(q,p,k) &= k^{3+\delta} F(\delta) \nn \\
 &= k^{3}(F_0\log(k/\mu)+\Lambda) + \mo(\delta)  \,,
\label{eq:dimregschem}
\end{align}
where $F(\delta)$ is a dimensionless constant, with $\delta$ expansion
\beq
F(\delta) = \frac{F_0}{\delta} + F_1 + \ldots
\eeq
In equation (\ref{eq:dimregschem}), $\Lambda$  is an infinite constant that we can subtract by counterterms.  Since we only need the $\log$ coefficient, $F_0$, we do not need to fully evaluate the divergent momentum integrals in equation (\ref{eq:aftereta}).

The identity
\beq
\int\,d^3p\int\,d^3q\,\delta^3(\mathbf{q}+\mathbf{p}+\mathbf{k})\,f(q,p,k)=\frac{2\pi}{k}\int_{0}^{\infty}\,dp\,p\int_{|p-k|}^{p+k}\,dq\, q\, f(q,p,k)
\label{eq:trickid}
\eeq
makes the extraction of $F_0$ efficient.  We see from the last line of equation (\ref{eq:dimregschem}) that multiplying the RHS by $k$ and taking 5 $k$ derivatives gives $24 F_0/k$.  Doing the same to the LHS, but swapping in the identity in equation (\ref{eq:trickid}), we find
\begin{equation}
F_0 = -\frac{\pi}{15}\,.
\label{eq:scalarf}
\end{equation}
At this point, the calculation joins that of \cite{delRio:2018vrj,Senatore:2009cf}.  As noted in section \ref{subsec:ccs}, the $\log(k/\mu)$ in equation (\ref{eq:dimregschem}) breaks dS isometries, and is clearly problematic as the ratio of a comoving to a physical momentum.  The resolution, originally found in \cite{Senatore:2009cf}, is to consistently deform all dependence on the number of spatial dimensions to $3+\delta$, including the index of the Hankel function in the graviton mode function ({\it cf.}~equation \eqref{eq:gravmode}).  Expanding in $\delta$ gives
\beq
\gamma^{s}_\mathbf{k}(\eta) = \frac{iH}{M_{\rm pl}\sqrt{2 \mu^\delta k^3}}(1+ik\eta)e^{-ik\eta}
\left[ 1 + \frac \delta 2 \log(-H\eta) + \frac \delta 2 u(-k \eta) + \ldots \right],
\label{eq:gravexp}
\eeq
where $u(-k\eta)$ is a combination of special functions whose details do not affect the $\log$ correction to the tensor power spectrum.  Readers interested in the full details should consult Appendix B of \cite{delRio:2018vrj}.

There is an important technical point worth elaborating here, though.  With the addition of the $\log$ term in equation (\ref{eq:gravexp}), doing the $\eta^\prime$ vertex integrals will introduce a factor of $\log(H/k)$.  This is because the time integrand dies off with a positive power of each vertex time as $\eta^\prime \rightarrow 0$, and is also suppressed at large $\eta^\prime$ by the Fourier factors, $e^{\pm i k \eta^\prime}$.  Thus, the contribution proportional to $\log$ term is an approximate $\delta$-function for $-\eta^\prime = 1/k$, and the deviation from this is just given by a logarithmically sensitive constant, which can be absorbed into the $\Lambda$ factor in equation \eqref{eq:dimregschem}.   The vanishing of the integrand at late times is guaranteed by a theorem in \cite{Weinberg:2005vy}, but it is obscured by our use of the anti-Euclidean-time-ordered formalism (equation (\ref{1loop}); it is also far from obvious in the traditional in-in calculation of \cite{delRio:2018vrj}).  This is because the limit $\eta_{1,2} \rightarrow 0$ of our imaginary time components does not actually correspond to a late-time interaction.  To make the late behavior of the integrand manifest, it is simplest to work with the nested commutator in-in formalism, stated in \cite{Weinberg:2005vy} and proven as a valid approach in \cite{Kaya:2018jdo,Baumgart:2020oby}.  It is straightforward to show that in the commutator basis for in-in propagators, and real vertex times, the integrand is suppressed at early times by at least four powers of the vertex times, justifying our treatment of the integrand as an approximate $\delta$-function giving $\log(H/k) + \text{const.}$

As discussed in section \ref{subsec:ccs}, while the $\log(H/\mu)$ correction is consistent and correct, it is insufficient to provide an observable effect.  For this, we need to further consider the deformation of dS to slow-roll inflation.  This follows just as in \cite{delRio:2018vrj,Senatore:2009cf}, which we refer to for details.  The net result is to adjust the log further
\beq
\log(H/\mu) \rightarrow -\eps_* \log(k/k_*),
\eeq
where $k_*$ is the pivot scale of the experiment of interest, and $\eps_*$ is the slow-roll parameter determined when it crossed the horizon.  Putting this all together gives a deformation of the tensor fluctuation power spectrum from a single conformally-coupled scalar, and by the argument of section \ref{sec:cftgrav}, an arbitrary extra CFT, with $c_{\rm extra}/c_{\rm scalar} = N_{\rm CFT}$,
\beq
\Delta \mathcal{P}_\gamma = +\eps_* \frac{2H_k^2}{\pi^2 M_{\rm pl}^2} \frac{N_{\rm CFT}}{16\pi^2} \frac{H_*^2}{M_{\rm pl}^2} \left[\frac{1}{15}\log \left( \frac{k}{k_*} \right) \right]\,.
\label{eq:explcorrrecap}
\eeq
As a nontrivial check on our CCS result, we see that the modification to the tilt, $\Delta n_t$, determined from this $\Delta \mathcal{P}_\gamma$ (equation \eqref{eq:tiltcorr}) exactly matches that obtained directly from the graviton one-loop effective action, equation (\ref{eq:nteff}).

\subsection{Fermion}

To test equation (\ref{eq:explcorrrecap}) and clear up confusion in the literature, we explicitly compute the one-loop effect of coupling a massless Dirac fermion to gravity in an inflationary background.  The calculation is extremely similar to the CCS case, but we explicitly show the initial stages.
In pure dS, the fermion mode functions are
\begin{align}
X_{\mathbf{k},\lambda}(\eta)=\frac{e^{-ik\eta}}{a^{3/2}(\eta)}u_{\mathbf{k},\lambda}\\
W_{\mathbf{k},\lambda}(\eta)=\frac{e^{ik\eta}}{a^{3/2}(\eta)}v_{\mathbf{k},\lambda}
\end{align}
where $\lambda$ is a label for helicity. We normalize our spinors such that
\begin{align}
\sum_{\lambda}\bar{u}_{\mathbf{k},\lambda}u_{\mathbf{k},\lambda}=\sum_{\lambda}\bar{v}_{\mathbf{k},\lambda}v_{\mathbf{k},\lambda}=\frac{\gamma^{\mu}k_\mu}{2k}\,,
\end{align}
with $k_\mu\equiv(|\mathbf{k}|,\mathbf{k})$. Therefore, from the interaction in equation (\ref{eq:fermint}) the loop correction to the graviton propagator is
\begin{align}
\left \langle \gamma^{\mathbf{k}\,s}_{ij} (\eta) \gamma^{\mathbf{-k}\,s^\prime}_{ij} (\eta) \right \rangle_2 &=
-\frac{1}{32}\int_{-\infty}^{\infty}\,d\eta_1 a^3(\eta+i\eta_1)\int_{-\infty}^{\infty}\,d\eta_2\,a^3(\eta+i\eta_2)\int\,d^3x_1\int\,d^3x_2 \nn \\
&\quad \times \langle\gamma_{ij}(x,\eta)\gamma_{ij}(y,\eta)\gamma_{ab}(x_1,\eta+i\eta_1)\gamma_{cd}(x_2,\eta+i\eta_2)\rangle \nn \\
&\quad\times \left[\langle\bar\psi_1\gamma_{(a}\partial_{b)}\psi_1\bar\psi_2\gamma_{(c}\partial_{d)}\psi_2\rangle -\langle\bar\psi_1\gamma_{(a}\partial_{b)}\psi_1\partial_{(c}\bar\psi_2\gamma_{d)}\psi_2\rangle \right. \nn \\
&\left. \qquad -\langle\partial_{(a}\bar\psi_1\gamma_{b)}\psi_1\bar\psi_2\gamma_{(c}\partial_{d)}\psi_2\rangle+\langle\partial_{(a}\bar\psi_1\gamma_{b)}\psi_1\partial_{(c}\bar\psi_2\gamma_{d)}\psi_2\rangle \right],
\end{align}
where we have written $\psi(x_i,\eta+i\eta_i)$ as $\psi_i$ for brevity, and suppressed $\overline T_E$ in the expectation values. Since the graviton polarization tensors are symmetric, we can ignore the symmetrization of the indicies in the fermion 4-point functions. In addition, the transverse property of the graviton polarization allows us to write each of the four fermion 4-point functions in the same form.  We have
\begin{align}
\left \langle \gamma^{\mathbf{k}\,s}_{ij} (\eta) \gamma^{\mathbf{-k}\,s^\prime}_{ij} (\eta) \right \rangle_2 &= -2\, \delta^{s s^\prime} \int_{-\infty}^{\infty}\,d\eta_1 \int_{-\infty}^{\infty}d\eta_2\int d^3p\,\int\,d^3q\,\delta^3(\mathbf{q}+\mathbf{p}+\mathbf{k})\left(\frac{H^2}{M_{\rm pl}{}^2}\right)^2\epsilon_{ab}^{*\, s}(\mathbf{k}) \, \epsilon_{cd}^{s^\prime}(\mathbf{k}) \nn \\
&\quad \times \frac{e^{-k|\eta_1|}}{2k^3}\frac{e^{-k|\eta_2|}}{2k^3}(1+k^2\eta^2+ik^2\eta\eta_1+k|\eta_1|)(1+k^2\eta^2+ik^2\eta\eta_2+k|\eta_2|) \nn \\
&\quad \times\left[\frac{e^{(p+q)|\eta_1-\eta_2|}}{pq}(2p_ap_bp_cp_d-p_\mu q^\mu\delta_{ac}p_bp_d)\right]
\end{align}
We compute the time integrals just as in section \ref{subsec:ccs} and use the fact that $\epsilon_{ab}(k)\epsilon^*_{cd}(k)\delta_{ac}p_bp_d=2p^2\sin\theta$, getting
\begin{align}
\left \langle \gamma^{\mathbf{k}\,s}_{ij} (\eta) \gamma^{\mathbf{-k}\,s^\prime}_{ij} (\eta) \right \rangle_2 &= - \frac 1 2 \left(\frac{H^2}{M_{\rm pl}^2}\right)^2\frac{1}{k^6}\int d^3p\,\int\,d^3q\,\delta^3(\mathbf{q}+\mathbf{p}+\mathbf{k})\nn \\
&\quad \times \frac{16k^3+29k^2(q+p)+20k(q+p)^2+5(q+p)^3 }{k\, (q+p+k)^4} \nn \\
&\quad \times \left[\frac{1}{pq} \left(p^4\sin^4\theta-p^3 q\sin^2\theta-p^4\sin^2\theta-p^3k\cos\theta\sin^2\theta \right) \right]
\end{align}
Once again, we can write the trigonometric functions in terms of $q$, $p$, and $k$ using $q^2=p^2+k^2+2pk\cos\theta$ and compute the momentum integrals as we did for the scalar case, defining $f(q,p,k)$ as everything to the right of the $\delta$-function.  Then, by analogy with equation (\ref{eq:dimregschem}), this gives the log coefficient
\begin{equation}
F_0 = +\frac{4\pi}{5}\,.
\end{equation}
To get the factor that enters the power spectrum (the analog of the $1/15$ in equation \eqref{eq:explcorrrecap}), we first need to combine $F_0 = 4\pi/5$ with the overall $-1/2$.  We see from equation (\ref{eq:tfpsdef}) how to get the contribution to $\Delta \mathcal{P}_\gamma$.   We sum over polarizations, getting a factor of 2, and multiply by $k^3/(16\pi^5)$.  To compare with equation (\ref{eq:explcorrrecap}), we then pull out $1/8\pi^4$ in front and multiply by $-1$, since introducing slow-roll converts $\log(H/\mu) \rightarrow -\eps_* \log(k/k*)$.  Taken all together, this leaves us with a factor of $(2/5)$, which is exactly $+6\times$ the $(1/15)$ in equation \eqref{eq:explcorrrecap}.  Thus, the correction to the tensor fluctuation power spectrum, $\Delta \mathcal{P}_\gamma$, from a Dirac fermion matches the result from conformal symmetry, equation (\ref{eq:ccharges}).

\bibliography{arxiv3}
\bibliographystyle{JHEP}

\end{document}